\documentclass[10pt,letterpaper,twocolumn]{article} 

\usepackage{ol2}
\usepackage{hyperref}
\usepackage{breakurl}
\usepackage{amsmath}
\usepackage{graphicx}
\usepackage{amssymb}
\newcommand{\I}{\mathrm{i}}
\newcommand{\D}{\mathrm{d}}
\newcommand{\E}{\mathrm{e}}
\begin{document}
\twocolumn[ 
\title{Proof-of-concept implementation of the massively parallel algorithm for
simulation of   dispersion-managed
WDM optical fiber systems}
\author{Alexander~O.~Korotkevich$^{1,2}$ and Pavel~M.~Lushnikov$^{1}$}
\address{$^{1}$ Department of Mathematics and Statistics,
University of New Mexico, Albuquerque, NM 87131, USA\\
$^{2}$ L.\,D.~Landau Institute for Theoretical Physics, Kosygin St. 2,
Moscow, 119334 Russian Federation}
\email{alexkor@math.unm.edu}


\begin{abstract}
We perform a proof-of-concept implementation of the massively parallel algorithm  (P.M. Lushnikov, Opt. Lett., v. 27, 939 (2002))        for
simulation of  dispersion-managed
wavelength-division-multiplexed optical fiber systems. Linear scalability of the algorithm with the number of computer cores is demonstrated.
Exact result on the accuracy of the implemented algorithm is found analytically and confirmed numerically as well as it is compared with the accuracy of the standard  split-step algorithm.
\end{abstract}
\maketitle
]

A wavelength-division-multiplexed (WDM) dispersion-managed (DM) optical fiber system is the basis of current high-bit-rate optical communications.
Next generation of these systems will use both the amplitude and phase of the optical signal  as a carrier of information (see e.g. \cite{RenaudierEtAlJLightwaveTech2008,RadicNatPhot2010})
to achieve higher system performance.
WDM systems are weakly nonlinear ones with a linear dispersion length typically in rage of tens while a nonlinear length is at several hundreds of kilometers~\cite{LushnikovOL2000,LushnikovOL2001,GabitovLushnikovOL2002,GabitovIndikLushnikovMollenauerShkarayevOL2007}.
Nonlinearity is a  major factor limiting performance of such systems while linear effects can be significantly compensated by coherent detection.

WDM requires propagation of a wide range of frequnces through optical fiber coupled by the nonlinearity.
Path-averaged group-velocity dispersion (GVD) effects cause optical pulses in distinct WDM
channels to move with different group velocities. Consequently modeling of WDM systems requires
simulating a long time interval, which determines needed high resolution in the frequencies.
All that makes accurate numerical simulations enormously challenging with very large number of Fourier modes $N$ needed to be resolved.
The standard
algorithm for such simulation is an operator splitting, or split-step algorithm (SS). It involves several Fourier transforms for every spatial step along optical fiber.
The fast Fourier transform (FFT) algorithm computes each such transformation in $O(N\log N)$ operations of
multiplication. The efficiency of using of supercomputers for SS is limited because parallel algorithms for one dimensional FFT (contrary to multidimensional FFT) provide
only very moderate speed up. For example, one of the leading implementations~\cite{MKLBenchmark}
shows at best four times acceleration on 16 processor cores on the system with shared memory. Further increase of CPU number appears to be inefficient.
Further increase of number of processors on conventional systems requires use of a distributed memory approach
(cluster), which has higher latency of nodes interconnection media. The experimental data shows~\cite{MKLBenchmark} that at number of harmonics up to $2^{21}$, a shared memory
approach is more efficient. At a higher number of harmonics, moderate acceleration can be achieved on a cluster,
although scaling would still be far from linear.

Here we demonstrate the proof-of-concept realization of the massively parallel algorithm (MPA) for simulation of WDM systems that is free from all these limitations.
 MPA was proposed by one of the authors of this Letter in~\cite{LushnikovOL2002} and exploits weak nonlinearity of WDM system.
We demonstrate 
the linear
 scalability of performance with number of computer cores.
We also obtain the
exact result on the accuracy of the algorithm in comparison with SS.
The results are in full agreement with numerics. 

We neglect polarization effects, 
stimulated Raman scattering, and Brillouin scattering. Then the
propagation of WDM optical pulses in DM fiber is described by a scalar nonlinear Schr\"odinger
equation
\begin{eqnarray}
\I A_z -\frac{1}{2} \beta_2(z)  A_{tt}-\frac{\I}{6}\beta_3(z)A_{ttt}   + \sigma(z) |A|^{2} A
= \I G(z)A, \label{nls1}
\end{eqnarray}
where $ G(z)\equiv\{-\gamma +[\exp{(z_a\gamma)}-1]\Sigma^N_{k=1}\delta(z-z_k)\}$; $z$ is the
propagation distance along an optical fiber; $A(t,z)$ is the slow amplitude of light; $\beta_2$
and $\beta_3$ are the first- and second-order GVD, respectively, which are periodic functions of $z$;
$\sigma=(2\pi n_2)/(\lambda_0 A_{eff})$ is the nonlinear coefficient; $n_2$ is the nonlinear
refractive index; $\lambda_0$ is the carrier wavelength; $A_{eff}$ is the effective
fiber area; $z_k=kz_a\ (k=1, 2,\ldots, N)$ are amplifier locations; 
and $\gamma$ is the loss coefficient. Distributed amplification can be also included in $G(z).$

Applying Fourier transform $\hat A(\omega,z)=\widehat F[A(t,z)]=\int^\infty_{-\infty}A(t,z)\exp[{\I \omega t}]\D t$
to Eq.~$(\ref{nls1})$, using the change of variables $\hat A  (\omega, z)\equiv \hat \psi (\omega, z)\exp[{\I\beta(\omega, z) +\int^z_{z_0} G(z')dz'}]$
and integration over $z$ result in the integral equation
\begin{eqnarray}
\label{integral_nlse}
\hat \psi (\omega, z) = \hat \psi (\omega, z_0) + \I \int_{z_0}^{z} \sigma(z') \widehat F [|A(t,z')|^2 A(t,z')]\nonumber \\
\times \E^{-\I\beta(\omega, z')-\int^{z'}_{z_0} G(z'')dz''}\D z',
\end{eqnarray}
where
$\beta(\omega, z) = \int_{z_0}^{z}\big[\frac{\omega^2}{2}\beta_2(z') + \frac{\omega^3}{6}\beta_3(z')\big] \D z'$.

Case $\hat \psi (\omega, z) =const$ corresponds to the exact solution of the linear part of Eq.~$(\ref{nls1})$ [or, equivalently,  setting $ \hat F [\cdot]\equiv 0$ in  $(\ref{integral_nlse})$].
Assume that the nonlinearity is weak, $z_{nl}\gg z_{disp}$, where
 $z_{nl}\equiv 1/|p|^2$ is a characteristic nonlinear length,
 $z_{disp}\equiv\tau^2/|\beta_2|$ is the dispersion length, and $p$ and $\tau$ are  typical pulse amplitude and
width, respectively. Then  $\hat \psi(\omega, z)$ is a slow function of $z$ on
any scale $L\ll z_{nl}$
(see~\cite{GT1996a,GT1996b,LushnikovOL2001}).
We solve Eq.~$(\ref{nls1})$ by iterations for $0\le z-z_0\le L$, where we have a freedom of choice of $L$
with the only condition that $L\lesssim z_{disp}.$ For the first iteration we set $\hat \psi^{(0)}(\omega, z)=\hat \psi(\omega, z_0)=\hat A  (\omega, z_0)$ and, respectively,
 $A(z,\omega)=\hat \psi^{(0)}(\omega, z)\E^{\I\beta(\omega, z) +\int^z_{z_0} G(z')dz'}$  on the right hand side (rhs) of Eq.~$(\ref{nls1})$,
which gives the first iteration $\hat \psi^{(1)}(\omega, z)$
for the left-hand side (lhs) of Eq.~$(\ref{nls1})$. Similarly, substitution of $\hat \psi^{(n-1)}(\omega, z)$ in the rhs. of Eq.~$(\ref{nls1})$ gives  $\hat \psi^{(n)}(\omega, z)$ in the lhs of Eq.~$(\ref{nls1})$ for $n=1,2,\ldots$.

In simulations we use  $\hat \psi(\omega, z_0)$ with $z_0=mL$ for a given $m=0,1,\ldots$ to perform a total number of iterations $n_{tot}$ to approximate $\hat \psi(\omega, z_0+L)$ as
$\hat \psi^{(n_{tot})}(\omega, z_0+L)$. Then we
use that approximate value as starter for the next spatial interval by setting $z_0=(m+1)L$
and proceeding in a similar way.

Assume that the interval $z_0\le z\le z_0+L$
includes $M+1$ equally spaced points $z_0, z_1,\ldots , z_M=z_0+L$.
The MPA is based on these iterations as follows:\\
{\bf 1.}
$\hat A (\omega, z_0) = \hat \psi (\omega, z_0)$, copy $\hat \psi (\omega, z_0)$ in
$\hat \psi (\omega, z)$ at all  $z$.\\
{\bf 2.}
Find $\hat A (\omega, z) = \hat \psi (\omega, z) \E^{\I\beta(\omega, z)+\int^z_{z_0} G(z')dz'}$ at all $z$.\\
{\bf 3.}
In order to return to $t$-domain, calculate independent Fourier transforms
$A(t, z) = \hat F^{-1} [\hat A(\omega, z)]$ at all  $z$.\\
{\bf 4.}
Calculate independent Fourier transforms $\hat V(\omega, z) = \hat F [|A(t, z)|^2 A(t, z)]$ at all $z$.\\
{\bf 5.}
Numerical integration (summation) by trapezoidal rule of the integral in Eq.~(\ref{integral_nlse}) using $\hat V(\omega, z)$ from step {\bf 4}.  Save intermediate results of integration at every $z$.\\
{\bf 6.}
For the second, third, etc., iterations go to step {\bf 2}.\\
{\bf 7.}
Reconstruct $\hat A (\omega, z_M)$ on the far edge of interval.

\begin{figure}[hbt]
\centering
\includegraphics[width=3.0in]{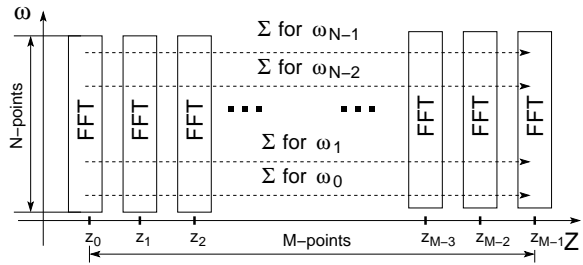}
\caption{\label{AlgoScheme} Schematic representation of MPA.}
\end{figure}

The MPA is schematically shown in Fig. \ref{AlgoScheme}.
All  steps of the MPA are computed parallelly.
The most time-consuming steps are {\bf 3} and {\bf 4}. All  FFTs at each $z$
are independently performed in  CPU cores (vertical bars in Fig.~\ref{AlgoScheme}).
Calculations in steps {\bf 2} and {\bf 5} are done for every harmonic independently
(dashed horizontal line in Fig.~\ref{AlgoScheme}).




We implement the MPA for shared memory symmetric multiprocessor (SMP) architecture.
The only powerful SMP computer in exclusive use was
the HP SuperDome 64000 supercomputer, equipped with 64 HP PA-RISC
processors
({\tt http://jscc.ru}).
Both processors and memory bandwidth
are outdated and relatively slow. However, for the proof-of-concept simulation
the main criteria is the number of processors in the system.

\begin{figure}[hbt]
\centering
\includegraphics[width=3.0in]{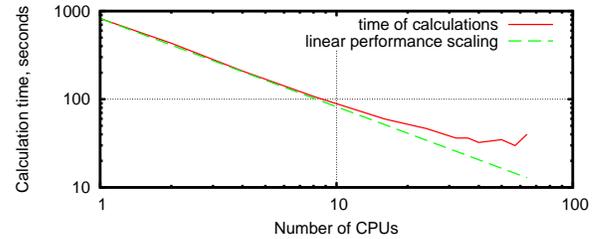}
\caption{\label{Scalability} Scalability of MPA on HP SuperDome 64000.}
\end{figure}

Simulations were performed in a setup identical to the one used in~\cite{LushnikovOL2002} with pseudorandom sequences of optical pulses in five channels of 20 periods of
WDM DM system. The main difference between the current MPA implementation and the original algorithm \cite{LushnikovOL2002} is in a more efficient way to handle summation in step {\bf 5},
optimizing CPUs cache use.
 We achieve $\sim 30$ times speed up with respect to a single processor version
of the code.
Fig.~\ref{Scalability} shows the scalability of performance.
Scaling is close to linear up to 32 processors. Then memory bandwidth limitation
of the available SMP computer makes further parallelization less efficient. Another reason for that
was restriction on the memory usage, which limited the number of Fourier harmonics.
However, we see a clear tendency of the scalability improvement with an increase of the number of harmonics because the longer time computer spends in FFTs the less important communications are.



To find the accuracy of the MPA we put the exact solution of Eq.~(\ref{nls1}) in the operator form as
\begin{equation}\label{operatorLN}
A_{exact}(z)=\exp [ i(\hat {\cal L}+\hat {\cal N})z]A(0),
\end{equation}
where $\hat {\cal L}$ represents all linear terms and $\hat {\cal N}$ represents the nonlinear term in Eq.~(\ref{nls1}) and we set $z_0=0$.
Here and below for brevity we omit the argument $t$ of function $A(z,t)$. We assume
that $\beta_2(z), \ \beta_3(z), \sigma(z), \ G(z)$ are constant functions of $z$ at each interval of length $L$, although a generalization to more general dependence on $z$ is straightforward.
SS uses the efficiency and high precision of the simulations for   $\exp  [ i\hat {\cal L}z ]A(0)$ and $\exp  [ i\hat {\cal N}z ]A(0)$.
But $\hat {\cal L}$ and $\hat {\cal N}$ do not commute and we need to approximate Eq.~(\ref{operatorLN}) for $z=L$ by the composite $M$ steps of SS as
\begin{equation}\label{SplitStepComposite}
A_{SS}(L)\equiv\exp  [ \I\hat {\cal L}\Delta z/2  ]\hat Q^M\exp  [ -\I\hat {\cal L}\Delta z/2  ]A(0),
\end{equation}
where $\Delta z\equiv L/M$,  $\hat Q\equiv \exp [ \I\hat {\cal N}\Delta z  ]\exp [ \I\hat {\cal L}\Delta z  ]$. Taylor series expansion of operators in Eqs.~(\ref{SplitStepComposite}) and
(\ref{operatorLN}) gives the following error  $r_1 \equiv  A_{exact}(L)-A_{SS}(L)$ of composite SS for arbitrary $M$ assuming $L\lesssim z_{disp}\ll z_{nl}$:
\begin{equation}\label{psiSScompositeerror}
    r_1=i[L^3/(2M^2)]P_L+ [L^3/M^2]O(\hat {\cal L}\hat {\cal N}^2 A_0),
\end{equation}
where $A_0\equiv A(0)$,    $O(\hat {\cal L}^k \hat {\cal N}^l A_0)$ means different combinations of terms with $k$th power of $\hat {\cal L}$ and $l$th power of $\hat {\cal N}$. Also,
$P_L=\frac{1}{6}\bar A_0(\hat {\cal L}A_0 )^2-\frac{1}{3}A_0|\hat {\cal L}A_0 |^2
    +\frac{1}{6}| A_0|^2\hat {\cal L}^2A_0-\frac{1}{3}\hat {\cal L}(|A_0|^2\hat {\cal L}A_0 )
    +\frac{1}{12} A_0^2\hat {\cal L}^2\bar A_0+\frac{1}{6}\hat {\cal L}(A_0^2\hat {\cal L}\bar A_0 )
    +\frac{1}{12}\hat {\cal L}^2(|A_0|^2A_0 )$ represents all terms with the second power in $\hat {\cal L}.$
Note that the operator expansions for arbitrary $M$ are not trivial and requires the extensive use of the symbolic computations.

Discretization of iterations over $z$ in Eq.~(\ref{integral_nlse}) with $
\hat \psi_j(\omega)\equiv \hat \psi(z_j,\omega), \quad z_j=j\Delta z,  \quad j=0,1,\ldots, M$
 at each $N$ discrete values of $\omega$ is given by:
\begin{equation}\label{psinplus1}
  \hat \psi^{(n+1)}_l(\omega)= \hat \psi^{(n)}_0(\omega)+\I\frac{V^{(n)}_0}{2}+\I\sum\limits^{l-1}_{j=1}V^{(n)}_j+\I\frac{V^{(n)}_l}{2},
\end{equation}
where  $l=1,2,\ldots,M,\; V^{(n)}_j\equiv \sigma(z_j)\hat F[ |A^{(n)}_j|^2 A^{(n)}_j]$\\
$\Delta z\exp[-\I\beta(\omega, z_j)-\int^{z_j}_{0} G(z)dz],\;\hat A^{(n)}_j=\hat\psi^{(n)}_j\exp[\I\beta(\omega, z_j)+\int^{z_j}_{0} G(z)dz]$,
and $ \hat\psi^{(n)}_l(\omega)$ is the $n$th iteration of $\psi$ while for zero iteration $ \psi^{(0)}_j=A_0, \ j=0,1,2,\ldots, M$.
From comparison of the operator expansion for $n$th iteration with the operator expansion of the exact solution Eq.~(\ref{operatorLN}) we obtain
 (again assuming that  $\beta_2(z), \beta_3(z), \sigma(z), G(z)$ are constant functions of $z$) that the error
 $r_2 \equiv  A_{exact}(L)-A^{(n)}(L)$ of composite SS for arbitrary $M:$
\begin{equation}\label{psinerror}
   r_2=i\frac{L^3}{M^2}P_L+ \frac{L^3}{M^2}O\left(\hat {\cal L}\hat {\cal N}^2 A_0\right )  +O\left( \hat {\cal N}^{n+1}A_0\right ),
\end{equation}
where $P_L$ is the same as in Eq.~(\ref{psiSScompositeerror})
and we assume that $n\ge 3$, which ensures that $r_2$ at leading order $O(L^3)$ does not depend on $n$. For $n=2$ the additional error term is $O([L^4/M^2]\hat {\cal L}^3\hat {\cal N} A_0)$,
which can be of the same order as $L^3P_L/{M^2}$ provided $L\sim z_{disp}.$ But for practical realization of MPA we expect  that $L\ll  z_{disp}$ and then $n_{tot}=2$ can be also an optimal choice.

Error term $\propto P_L$ dominates in Eq.~(\ref{psiSScompositeerror}) while in Eq.~(\ref{psinerror}) it competes with the last term in the rhs which has the order $O (\hat {\cal N}^{n+1}A_0) \sim (L/z_{nl})^{n+1} A_0$ and is independent
of $M$ because it results from the iterations of Eq.~(\ref{integral_nlse}) in the continuous limit $M\to \infty$.
An increase of $n$ ensures dominance of $\propto P_L$ in Eq.~(\ref{psinerror}) because $L\lesssim z_{disp}\ll z_{nl}.$
Then we conclude from comparison of Eqs.~(\ref{psinerror}) and (\ref{psiSScompositeerror}) that SS error is twice smaller than the MPA. So to  match the  accuracy of SS it is enough for
the MPA to take $M$ by a factor $2^{1/2}$ larger.
Respectively, the MPA requires  a minimum $2^{1/2}n+1$ CPU cores to outperform SS ($2^{1/2}n$ would be the exact match of performance).
For example, in simulations with the parameters of Fig. \ref{Scalability}, 250 DM periods and $n=2$ we obtained   the ratio $9.2$ of SS and MPA computation times  at equal accuracy and 32 cores, which is close
to the theoretical prediction  $2^{-1/2}32 n^{-1}$.

\begin{figure}[hbt]
\centering
\includegraphics[width=3.0in]{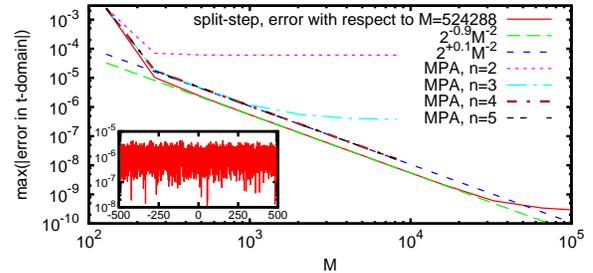}
\caption{\label{Accuracy} Errors in $L_{\infty}$ norm (maximum over $t$) of MPA and SS vs. $M$. Results for 4th and 5th iterations
are visually indistinguishable with the theoretical scaling line. Inset shows the error (normalized to $\max|\psi(t)|$) with $t$ for 20-channel WDM system after $10^4$ km.}
\end{figure}

To check these analytical predictions  we simulated a three-channel WDM system
over one period of DM fiber system with $2^{11}$ frequency harmonics, $L=20$km for the standard fiber, and other parameters as in~\cite{LushnikovOL2002}. As a ``numerically exact'' we use SS
with $2^{21}$ grid steps over one DM system period.
 Fig.~\ref{Accuracy} shows
that the error of the MPA with $n=3$ scales as one for
 SS. That is, $n=3$ is enough to neglect $O( \hat {\cal N}^{n+1}A_0)$ term in Eq.~(\ref{psinerror}). Errors for MPA and SS are different by a factor 2 in full agreement with
Eqs.~(\ref{psiSScompositeerror}) and (\ref{psinerror}). We also compare the MPA and SS for simulation of the transoceanic distance $10^{4}$ km (250 DM periods) of the realistic
WDM system with 20 channels using
$N=2^{13}$, $M=2^{14}$, $L=1.25$km. That system has $(20/3)^2$ higher nonlinearity than above so we decreased $L$.
The inset of Fig.~\ref{Accuracy} shows the error of the MPA with $n=3$ in that case.
A ratio of MPA and SS $L_{\infty}$ errors (i.e., max over $t$ in that inset and similar for SS) is $\simeq 2.1$, again close to 2.

In conclusion, we have demonstrated the feasibility of the MPA. Scaling of the parallel version on the available SMP machine
was close to linear up to 32 processing threads even on outdated architecture and with a very restricted size
of FFT arrays.
We propose using a shared memory model for parallel computation,
which has lower penalties due to interprocess communications, and exploiting the power of modern graphics processing units (GPUs).
Nvidia Tesla C2070 GPU
has 448 cores and 6GB
memory, which appears quite suitable for the MPA.  




Work of A.\,K. was partially supported by grants: RFBR 09-01-00631-a, NSh-6885.2010.2,
and program ``Nonlinear Dynamics''.
P.\,L. was supported by the National Science Foundation grant PHY 1004118.

\end{document}